\documentclass[fleqn,usenatbib]{mnras}

\usepackage{newtxtext,newtxmath}
 
\usepackage[T1]{fontenc}

\DeclareRobustCommand{\VAN}[3]{#2}
\let\VANthebibliography\thebibliography
\def\thebibliography{\DeclareRobustCommand{\VAN}[3]{##3}\VANthebibliography}

\usepackage{graphicx}	
\usepackage{amsmath}	
\usepackage{amssymb}	

\title[Symmetry Energy Constraints from RSFs]{Resonant Shattering Flares as Multimessenger Probes of the Nuclear Symmetry Energy}

\author[D. Neill et al.]{
Duncan Neill,$^{1}$\thanks{E-mail: dn431@bath.ac.uk}
William G. Newton,$^{2}$
David Tsang$^{1}$
\\
$^{1}$Department of Physics, University of Bath, Claverton Down, Bath, BA1 1AL\\
$^{2}$Department of Physics and Astronomy, Texas A\&M University-Commerce, Commerce, TX, 75429-3011
}

\date{Accepted XXX. Received YYY; in original form ZZZ}

\pubyear{2015}

\begin{document}
\label{firstpage}
\pagerange{\pageref{firstpage}--\pageref{lastpage}}
\maketitle

\begin{abstract}
The behaviour of the nuclear symmetry energy near saturation density is important for our understanding of dense nuclear matter. This density dependence can be parameterised by the nuclear symmetry energy and its derivatives evaluated at nuclear saturation density. In this work we show that the core-crust interface mode of a neutron star is sensitive to these parameters, through the (density-weighted) shear-speed within the crust, which is in turn dependent on the symmetry energy profile of dense matter. We calculate the frequency at which the neutron star quadrupole ($\ell = 2$) crust-core interface mode must be driven by the tidal field of its binary partner to trigger a Resonant Shattering Flare (RSF). We demonstrate that coincident multimessenger timing of an RSF and gravitational wave chirp from a neutron star merger would enable us to place constraints on the symmetry energy parameters that are competitive with those from current nuclear experiments.
\end{abstract}

\begin{keywords}
dense matter -- stars: neutron -- stars: oscillations -- neutron star mergers -- gravitational waves -- equation of state
\end{keywords}



\section{Introduction}

Neutron stars contain the most extreme matter in the universe. They act as natural laboratories to investigate nuclear physics, allowing us to study the physics of matter at nuclear densities. To investigate the internal structure of these compact stars we must probe them using observational phenomena.

Short gamma-ray bursts (SGRBs) \citep{kouveliotou1993identification,d2015short} likely originate from the merging of binary neutron stars \citep{eichler1989nucleosynthesis,fong2009hubble}. These bursts are characterised by a large peak in the gamma-ray count that lasts for $\sim 2$ second or less. Around $\sim3-10$\% of SGRBs are preceded by a `precursor' flare \citep{zhong2019precursors,troja2010precursors}. 
Precursor flares can be identified as a lower, separate peak in the gamma-ray count a short time ($\sim 0.1-5.0$ s) before the main peak \citep{zhong2019precursors}. 
Resonant Shattering Flares \citep{tsang2012resonant} are relatively isotropic, short ($\sim 0.1$s duration) gamma-ray flares that are triggered by a tidal resonance of the binary, and can appear as either precursor flares, or orphan flares if the main SGRB is beamed away from the observer.

Recent observations of neutron star merger GW170817 \citep{abbott2017gw170817,goldstein2017ordinary} have begun a new era of multi-messenger astronomy involving gravitational waves and counterparts across the electromagnetic spectrum, allowing an unprecedented probe into the physics of binary neutron star mergers \citep[see e.g. ][ and references therein]{raithel2019constraints}. Here, we will show how multi-messenger coincident timing of a gravitational-wave chirp and the prompt-gamma ray emission from a Resonant Shattering Flare can be used to determine the frequency of a particular neutron star oscillation mode and hence constrain fundamental parameters in nuclear physics.

\subsection{Resonant Shattering Flares}
During the gravitational-wave induced inspiral of neutron star binaries, the normal modes of a neutron star can become excited by resonant tidal interactions with its binary partner. As the binary orbit shrinks, the frequency of the orbit (and hence the gravitational wave frequency) increases. When the orbital frequency sweeps through the appropriate resonance windows, the normal modes are excited, causing their oscillations to rapidly grow in magnitude \citep{tsang2012resonant, tsang2013shattering, lai1994resonant}.

If a resonant mode is sufficiently large to cause a deformation that exceeds the breaking strain in the neutron star crust, the crust will fracture, depositing seismic energy into the crust. One such mode is the quadrupole crust-core interface ($i$) mode. This mode is caused by the discontinuity in bulk material properties between the crust and core, and has a sufficient overlap with the tidal field \citep{tsang2012resonant} for the resonance to quickly deposit energy into the mode. The fractures and seismic waves continue to be driven until the crust reaches its elastic limit, causing it to shatter. This scatters the seismic waves to high frequencies that are able to couple to the star's magnetic field, depositing energy into the magnetosphere and sparking a pair-photon fireball. 
Multiple colliding shells may be emitted over the course of the resonance window ($t_{\rm res} \sim 0.1$ seconds), leading to a single non-thermal burst with a duration of the resonance time, and maximum luminosity determined by the surface magnetic field strength \citep{tsang2013shattering}.

For sufficiently strong surface magnetic field, these Resonant Shattering Flares (RSFs) can be detectable well beyond the Advanced LIGO horizon. Coincident timing of the RSF prompt emission and the gravitational-wave chirp provides a precise measurement of the resonant mode frequency. Therefore, by using a model of the crust and core with consistent nuclear physics to calculate the $i$-mode frequency, the parameters of that model could be restricted to ranges which results in mode frequencies that closely match the observed gravitational wave frequency during the RSF.  

In this paper, we will explore the implications of a coincident multi-messenger detection of an RSF on constraining fundamental nuclear physics parameters. In particular we will explore the relationship between the $i$-mode frequency, neutron star structure, and the nuclear symmetry energy parameters that determines the behaviour of matter near nuclear saturation density.

\subsection{Nuclear Symmetry Energy and Neutron Star Structure}
Normal mode frequencies are dependent on the neutron star equation of state (EOS) (the relationship between the energy density and pressure within the star) and the composition of the crust as a function of density. At low densities the EOS can be accurately calculated by using the properties of experimentally measured nuclei \citep{baym1971ground}. However, at the extreme densities found in neutron stars we must rely on nuclear models to calculate the EOS, with different models giving significantly different EOSs due to the relative lack of experimental data that probes neutron-rich matter. This uncertainty is conveniently captured by the nuclear symmetry energy $E_{\rm sym}$, which encodes the average energy per nucleon required to convert symmetric nuclear matter (number of protons $Z$ = number of neutrons $N$) into pure neutron matter ($Z$=0). At nuclear saturation density (the density of nucleons in a nucleus) the symmetry energy can be parameterised by the coefficients of its density expansion, the first three of which are: the magnitude of the symmetry energy at saturation density
\begin{align}
J=E_{\rm sym}(n_0),    
\label{eq:sym_J}
\end{align}
\noindent the slope of the symmetry energy
\begin{align}
L=3n_0\frac{\partial E_{\rm sym}(n_{\rm b})}{\partial n_{\rm b}}\biggr\rvert_{n_{\rm b}=n_0},  
\label{eq:sym_L}
\end{align}
\noindent and the curvature
\begin{align}
K_{\rm sym}=9n_0^2\frac{\partial^2 E_{\rm sym}(n_{\rm b})}{\partial n_{\rm b}^2}\biggr\rvert_{n_{\rm b}=n_0},
\label{eq:sym_K}
\end{align}
\noindent where $n_0\approx 0.16$ fm$^{-3}$ is nuclear saturation density and $n_{\rm b}$ is the baryon density.

The symmetry energy parameters have some simple physical implications. The magnitude of the symmetry energy $J$ controls the proton fraction at saturation density, and the slope $L$ correlates with the pressure at that density. $K_{\rm sym}$ controls the derivative of pressure, thus determining how the pressure changes as one moves away from saturation density, and plays an important role determining the stability of matter and its compressibility. Through these effects, neutron star structure and composition are sensitive to the symmetry energy and related nuclear observables \citep{Brown:2000aa, Horowitz:2001aa, Steiner:2008aa, Fattoyev:2018aa}, and therefore astrophysical observations provide information on $E_{\rm sym}$ that complement nuclear experiment. Constraining the symmetry energy has been a priority in the field of nuclear physics over the past two decades \citep{tsang2009constraints,Li:2014fj,Horowitz:2014aa}, and the burgeoning field of multimessenger astronomy provides exciting opportunities to synthesise astrophysical observation and nuclear experiment to learn more about dense matter \citep{Tsang:2019ab}. Crust properties and their observational manifestations are particularly sensitive to the symmetry energy \citep{Newton:2013aaa}, including shear waves in the dense regions of the inner crust \citep{steiner2009constraints,Sotani2012a-crustosc,Sotani2013b-crustosc,Newton:2011crustosc}. RSFs are therefore a strong candidate to probe the symmetry energy \citep{tsang2012resonant,Newton:2013aaa}.

\section{Parameterised Neutron Star Equation of State and Composition}

\subsection{The Nuclear model} \label{sec:nuclear_model}

A consistent nuclear physics description of a neutron star requires an underlying nuclear model. We use an extended Skyrme mean field model for the uniform matter EOS \citep{Holt:2018aa}. Three parameters of the model affect only the pure neutron matter EOS, and can be written in terms of $J$, $L$ and $K_{\rm sym}$ \citep{newton2020nuclear,Balliet:2020aa}. This way, a given set of values \{$J$,~$L$,~$K_{\rm sym}$\} can be converted into a set of Skyrme models (each of which labeled by the 3 symmetry energy parameters) and corresponding crust models and equations of state.

We explore the distributions of $i$-mode frequencies predicted for three different sets of symmetry energy parameters.

\begin{figure}
\centering
\includegraphics[width=0.45\textwidth,angle=0]{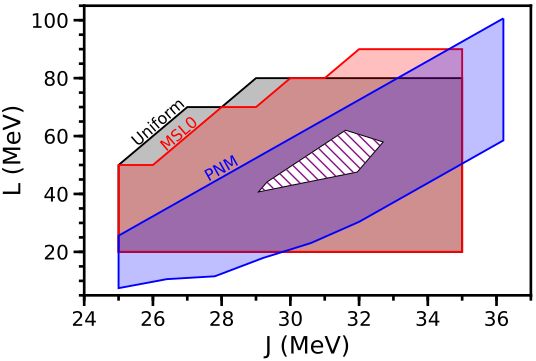}
\caption{The ranges used for the first two symmetry energy parameters $J$ and $L$. The ranges covered by the Uniform, MSL0 and PNM parameter sets are shown in grey, red and blue respectively. The Uniform and MSL0 sets are truncated in the high $L$, low $J$ region because stable neutron star crust models do not exist in that region of parameter space. Also pictured is the region in the intersection of a number of nuclear experimental \citep{Kortelainen:2010aa,chen2010density,Piekarewicz:2012aa,tsang2009constraints}, theoretical \citep{Hebeler:2010rz,Gandolfi:2012lr} and astrophysical \citep{Steiner:2010aa,Steiner:2013aa} constraints, as synthesized by \citet{lattimer2013constraining}.}
\label{fig:JLRanges}
\end{figure}

(i) Our most conservative range of EOSs uses symmetry energy parameters uniformly distributed over the ranges $25\leq J\leq 35$ MeV, $20\leq L\leq 80$ MeV, and $-200\leq K_{\rm sym}\leq 40$ MeV. These ranges were chosen to cover the ranges inferred from a variety of experimental probes \citep{liu2010nuclear,tsang2012constraints,lattimer2013constraining}. This will be referred to as our ``Uniform'' distribution, and is shown in the shaded blue region of Figure~\ref{fig:JLRanges} in $J$-$L$ space. Note that this set, as well as the MSL0 set below, are truncated in the high$L$-low$J$ region of parameter space as stable crust models do not exist for those combinations of parameters, as explained in section~\ref{sec:crust}.

(ii) The properties of pure neutron matter are the most important ingredient in the crust EOS, so relevant constraints for neutron stars comes from pure neutron matter theory. Great strides have been made in modeling pure neutron matter from first principles, particularly using chiral effective field theory. A useful way to parameterise these models is through a Taylor expansion of the Fermi liquid parameters that characterize the two-neutron interaction energy \cite{Holt:2017lr,Holt:2018aa}, three of which are shown to be sufficient to paramterise range of predictions for the PNM EOS from $ab-initio$ calculations. We translate this range into $J$, $L$ and $K_{\rm sym}$ space as detailed in \cite{newton2020nuclear} using a $9\times 9\times 9$ grid uniform in the Fermi liquid parameter space, and use this to form our second set of neutron star models. In $J$-$L$ space, this range is shown as the blue shaded region in Figure~\ref{fig:JLRanges}. This will be referred to as our ``PNM'' distribution.

(iii) In many Skyrme models, such as Sly4 \citep{chabanat1998skyrme} and SkI6 \citep{reinhard1995nuclear,nazarewicz1996structure}, there are only two parameters that control the PNM EOS alone, and therefore $K_{\rm sym}$ is not a free parameter. Likewise, only $J$ and $L$ are independent in many extractions of symmetry energy constraints. It will be useful to examine how important it is to take into account the $K_{\rm sym}$ degree of freedom, so for our third parameter set we emulate Skyrme models that have only $J$ and $L$ as free parameters. We take the dependence of $K_{\rm sym}$ from the MSL0 parameterisation of the Skyrme model \citep{chen2009higher}, which has previously been used to extract symmetry energy constraints from nuclear experiment. In that model, $K_{\rm sym}$ is related to $J$ and $L$ by \citep{newton2020nuclear}
\begin{align}
K_{\rm sym}=3.71L-11.13J+11.93\; \rm{MeV},
\end{align}
\noindent which restricts us to a single plane in the $J$,$L$,$K_{\rm sym}$ parameter space. We will refer to this as our ``MSL0'' distribution. Note that there is nothing physically special about this particular choice of relation between the symmetry energy parameters. 

We show the regions of $J$-$L$ space we sample for each of the three distributions in Figure~\ref{fig:JLRanges}. The regions we cover encompass a number of nuclear experimental constraints, the intersection of which is also shown as the white hatched region \cite{lattimer2013constraining}.  

\subsection{The crust model}\label{sec:crust}

To calculate the crust composition and EOS, we use our sets of extended Skyrme models in a compressible liquid drop model (CLDM) \citep{Newton:2013aa,Balliet:2020aa}. The model assumes a lattice consisting a single species of nucleus immersed in a neutron gas in a repeating unit cell (the Wigner-Seitz approximation). By minimising the energy of the unit cell with respect to the physical parameters of the cell - the neutron gas density $n_{\rm n}$, the cell radius $r_{\rm c}$, and the mass and charge number of the nuclear cluster $A,Z$ - we obtain the ground state composition and EOS of the crust at a given density. We can then calculate quantities required to model the normal modes in the crust, for example the shear modulus and frozen-composition adiabatic index as a function of baryon density $n_{\rm b}$ \citep{strohmayer1991shear,Chugunov:2010aa}. 

The shear modulus in the crust is given by 
\begin{align}
\mu=\frac{0.1106}{1+17810\left(\frac{ak_bT}{\left(Ze\right)^2}\right)^2}\frac{n_i\left(Ze\right)^2}{a},
\label{eq:mu_1991}
\end{align}
\noindent where $T$ is the temperature, $n_i$ is the ion density, $Z$ is the proton number of the nuclei, and 
\begin{align}
a=\left(\frac{3}{4\pi n_i}\right)^{\frac{1}{3}}.
\label{eq:mu_1991_a}
\end{align}

\noindent We conduct our calculations of the neutron star modes in the zero temperature limit. The ion number density can be written in terms of the fraction of nucleons in the neutron gas $X_{\rm n}$ through the relation \citep{Newton:2013aa}

\begin{equation}
    X_{\rm N} = 1 - X_{\rm n} = \frac{n_{\rm i}}{n_b}A,
    \label{eq:neutrons_ions}
\end{equation}

\noindent where $X_{\rm N}$ is the fraction of nucleons in the nucleus. 
We can therefore re-write the shear modulus as a function of $X_{\rm n}$ (at zero temperature) as \citep{Steiner:2008aa,Newton:2011crustosc}
\begin{equation}
	\mu = 0.1106 \left(\frac{4\pi}{3}\right)^{1/3} A^{-4/3} n_{\rm b}^{4/3} (1-X_{\rm n})^{4/3} (Ze)^2.
	\label{eq:altmu}
\end{equation}

The adiabatic index at constant composition is given by

\begin{equation} \label{eq:adiabatic_index}
\Gamma_{\rm 1} = \frac{n_{\rm b}}{P} \frac{dP}{dn_{\rm b}}\bigg|_{\rm constant \; composition} =  \frac{n_{\rm b}}{P} \bigg[\frac{dP_n}{dn_{\rm n}} +  x \frac{dP_e}{dn_{\rm e}}  \bigg],
\end{equation}

\noindent where $x$ is the average proton fraction and $P$ the total pressure. $P_{\rm n,e}$ are the pressures and $n_{\rm n,e}$ the number densities of dripped neutrons and electrons respectively.

We calculate crust models over the full range of each of our three symmetry energy ranges. However, EOSs with low values of $J$ and high values of $L$ do not result in stable crust models. Such EOSs have a symmetry energy that falls rapidly with decreasing density. If the magnitude of symmetry energy at saturation density is already small, then at sub-saturation densities the slope of the symmetry energy - and hence the pressure of pure neutron matter - must become very small, or even become negative. Since neutron matter pressure supports the inner crust in hydrostatic equilibrium, such crust models will be inherently unstable. This is the reason our ranges of $J$ and $L$ in Figure~\ref{fig:JLRanges} are truncated at high $L$, low $J$.

\subsection{The core model}\label{sec:core_model}

The Skyrme model is designed to describe nuclear interactions around nuclear saturation density. As one moves into the neutron star core, the increasing importance of relativistic effects, the possible appearance of hyperons at supersaturation density, and the likely transition from nucleonic to quark degrees of freedom in the inner core mean that the model is not suited to describe matter beyond about twice saturation density, and the symmetry energy loses its physical meaning. In order to explore the symmetry energy effects on RSFs, we do need to control the core EOS. We use the piecewise polytrope method \citep{Read:2009aa, Read:2009ab, Steiner:2010aa, Steiner:2013aa, Ozel:2009aa, Ozel:2010aa, Ozel:2016aa}: we fit two polytropes at supersaturation densities, one at a density of $n_1$=1.5$n_0$ and one at $n_2$=2.7$n_0$, as detailed in \cite{Newton:2018aa}. We then have three regions of the star: the crust and outer core, in which the pressure and energy density are given by the Skyrme EOS, and the two polytropic regions in which the pressures are given by

\begin{align}
\nonumber P = P_{\rm Skyrme} & \;\;\;\;\; n<n_1 \\
\nonumber P_1 = K_1 n^{\gamma_1}  & \;\;\;\;\;n_1 <n<n_2  \\
P_2 = K_2 n^{\gamma_2} & \;\;\;\;\;  n_2 < n
\end{align}

\noindent where continuity of pressure determines the constants $K_1$ and $K_2$. The energy density in the three density regions is obtained by integrating the first law of thermodynamics:

\begin{equation}
\epsilon_i = (1+a_i)n + \frac{K_i n^{\gamma_i}}{\gamma_1 -1}; \;\;\;\;\;\;\; a_i = \frac{\epsilon_{i-1} (n_i)}{n_i} - \frac{K_i n^{\gamma_i - 1}}{\gamma_i -1} - 1
\end{equation}

\noindent where $a_i$ are constants of integration, $i$=\{1,2\} and the subscript 0 labels the Skyrme EOS.

The speed of sound is 

\begin{equation}
\frac{c_{\rm s,i}(n)}{c} = \bigg( \frac{\gamma_i P}{P + \epsilon}\bigg)^{1/2}
\end{equation}

\noindent In the eventuality that the EOS becomes acausal ($c_{\rm s}>c$) at a given density $n_{\rm acausal}$, we transition to a causal EOS:

\begin{equation}
P_{\rm causal} = \epsilon = bn^{1/3} \;\;\;\;\;  n_{\rm acausal} < n
\end{equation}

\noindent where $\epsilon$ is the energy density, $b$ is a constant given by

\begin{equation} 
b = \frac{1+a}{n_{\rm acausal}} + K \frac{n_{\rm acausal}^{\gamma-2}}{\gamma-1}
\end{equation}

\noindent and $a$ is either $a_1$ or $a_2$ depending on which region the EOS becomes acausal in.

Each equation of state we generate is characterised by 5 parameters: the three symmetry energy coefficients $J$, $L$ and $K_{\rm sym}$ for the Skyrme-EOS, and the polytropic parameters $\gamma_1$ and $\gamma_2$. $\gamma_2$, which controls the high density part of the EOS, can be tuned to give a desired maximum mass. $\gamma_1$, which controls the EOS at intermediate densities in the core, can be tuned to give a particular moment of inertia of a 1.4$M_{\odot}$ star ($I_{1.4}$) while keeping the other parameters fixed. We can thus parameterise each EOS by $J$, $L$, $K_{\rm sym}$, $I_{1.4}$ and $M_{\rm max}$.

In this work we want to concentrate on how the crust models and their uncertainty affect the $i$-mode frequency, and so we fix the high density degrees of freedom $I_{1.4}$ and $M_{\rm max}$. As we shall see, the $i$-mode frequency is relatively insensitive to the stellar radius and therefore the high density EOS. We choose to fix $M_{\rm max}$ at 2.2$M_{\odot}$, comfortably above the maximum accurately measured pulsar mass \citep{cromartie2020relativistic} and consistent with maximum masses inferred from modeling of the binary neutron star merger resulting in GW170817. Given a value of $J$, $L$, $K_{\rm sym}$, and $M_{\rm max}$ the moment of inertia $I_{1.4}$ can be systematically varied between the minimum and maximum values allowed by causality. As we demonstrate in Figure~\ref{fig:I1.4}, this variation has only a small effect on the $i$-mode frequencies calculated, and so for each value of $J$, $L$, $K_{\rm sym}$, and $M_{\rm max}$ we choose the EOS model whose moment of inertia $I_{1.4}$ is the average of the maximum and minimum possible values of $I_{1.4}$ as the representative EOS.

\begin{figure}
\centering
\includegraphics[width=0.45\textwidth,angle=0]{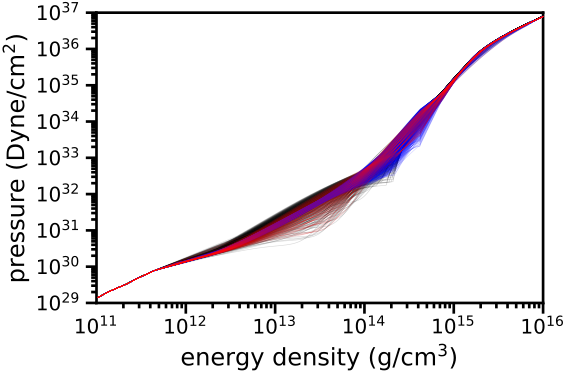}
\caption{The equations of state used in this work. The different colours are for the three different sets of symmetry energy parameter ranges we have used, with the black being for our Uniform ranges, blue for our PNM ranges, and red for our MSL0 ranges. Outside the density range shown here, the equations of state are indentical.}
\label{fig:EOSs}
\end{figure}

The resultant EOSs are shown in Figure \ref{fig:EOSs}. These are used with the Tolman-Oppenheimer-Volkoff (TOV) equations \citep{oppenheimer1939massive,tolman1939static} to determine the grid of stellar models for which the $i$-mode frequencies are calculated.

\section{Calculation of the Normal Modes}
We calculate the frequencies and radial/transverse displacements of the modes of a neutron star by linearly perturbing the equations defining the equilibrium state of neutron star matter. We assume that the binary lifetime is much larger than the spin-down and cooling times of the individual neutron stars, and so we may ignore rotation and high-temperature effects. We will also assume that the frequencies of the modes that we are interested in ($\sim 10 - 100$ Hz) result in oscillations that are significantly faster than the beta-equilibrium timescale. Therefore, weak-interactions do not have time to change the composition of a displaced mass element to more closely match the local composition.

\subsection{Basic Equations}
To construct a 2-component neutron star model (consisting of a solid crust and fluid core) we follow \citet{mcdermott1988nonradial}, beginning with the mass continuity equation, momentum conservation equation, and Poisson's equation:

\begin{align}
\frac{\partial\rho}{\partial t}+\nabla\cdot(\rho v)=0,
\label{eq:continuity_eqn}
\end{align}
\begin{align}
\frac{\partial v}{\partial t}=(v\cdot \nabla)v=\frac{1}{\rho}\nabla\cdot\sigma-\nabla\Phi,
\label{eq:momentum_eqn}
\end{align}
\begin{align}
\nabla^2\Phi=4\pi G\rho,
\label{eq:Poisson_eqn}
\end{align}
\noindent where $\rho$ is the energy density, $v$ is the velocity of the matter, $\sigma$ is the stress tensor, and $\Phi$ is the gravitational potential. By combining the linear perturbations of these equations, and taking the Cowling approximation by ignoring perturbations of the gravitational potential \citep{cowling1941non}, we obtain the wave equation
\begin{align}\nonumber
&&\omega^2u=-\nabla\left(\frac{\Gamma_1 p}{\rho}\nabla\cdot u\right)-\nabla\left(\frac{1}{\rho}u\cdot\nabla p\right)-\hat{r}A\frac{\Gamma_1 p}{\rho}\nabla\cdot u\\\nonumber
&&+\frac{1}{\rho}\biggr(\nabla\left(\frac{2}{3}\mu\nabla\cdot u\right)-\left(\nabla\mu\cdot\nabla\right)u-\nabla\left(u\cdot\nabla\mu\right)\\
&&+\left(u\cdot\nabla\right)\nabla\mu-\mu\left(\nabla^2 u+\nabla\left(\nabla\cdot u\right)\right)\biggr),
\label{eq:wave_eqn}
\end{align}
where $u(x,t)$ is the Lagrangian displacement,
\begin{align}
A=\frac{1}{\rho}\frac{d\rho}{dr}-\frac{1}{\Gamma_1P}\frac{dP}{dr}
\label{eq:schwartz_descrim}
\end{align}
\noindent is the Schwarzschild discriminant, and $\Gamma_1$ is the adiabatic index defined in equation~\eqref{eq:adiabatic_index}. The non-diagonal terms of the stress tensor are given by $\sigma_{ij} = \mu \nabla_i u_j$, assuming the isotropic shear modulus $\mu$ from equation~\eqref{eq:altmu}.

Taking the perturbations to have time dependence of the form $e^{i\omega t}$, where $\omega$ is the mode frequency, we have 
\begin{align}
u(x,t)=\xi(x)e^{i\omega t}.
\label{eq:time_seperation}
\end{align}
\noindent In spherical coordinates this can be further separated into radial and transverse components:
\begin{align}
\xi_r=U(r)Y_{lm},\;\;\;\xi_{\theta}=V(r)\frac{\partial Y_{lm}}{\partial\theta},\;\;\;\xi_{\phi}=\frac{V(r)}{\sin(\theta)}\frac{\partial Y_{lm}}{\partial\phi},
\label{eq:xi_seperation}
\end{align}
\noindent where $U(r)$ is the radial displacement, $V(r)$ is the transverse displacement, and $Y_{lm}$ are the spherical harmonics.

By using the separation of variables given in equations~\eqref{eq:time_seperation} and~\eqref{eq:xi_seperation}, the wave equation can be rewritten in terms of $U$ and $V$:\footnote{The version of equation~\eqref{eq:Ueqn} in \citet{mcdermott1988nonradial} has a typographical error.} 
\begin{align}\nonumber
&&\rho\omega^2U=\rho\frac{d\hat{\chi}}{dr}-A\Gamma_1 p\hat{\alpha}-\frac{d}{dr}\left(\frac{1}{3}\mu\hat{\alpha}\right)+\frac{d\mu}{dr}\left(\hat{\alpha}-2\frac{dU}{dr}\right)\\
&&-\mu\left(\frac{1}{r^2}\frac{d}{dr}\left( r^2\frac{dU}{dr}\right)-\frac{\ell(\ell+1)}{r^2}U+\frac{2\ell(\ell+1)}{r^2}V-\frac{2}{r^2}U\right),
\label{eq:Ueqn}
\end{align}
\begin{align}\nonumber
&&\rho\omega^2V=\rho\frac{\hat{\chi}}{r}-\frac{1}{3}\frac{\mu\hat{\alpha}}{r}-\frac{d\mu}{dr}\left(\frac{dV}{dr}-\frac{V}{r}+\frac{U}{r}\right)\\
&&-\mu\left(\frac{1}{r^2}\frac{d}{dr}\left(r^2\frac{dV}{dr}\right)-\frac{\ell(\ell+1)}{r^2}V+\frac{2}{r^2}U\right),
\label{eq:Veqn}
\end{align}
\noindent where:\footnote{The version of equation~\eqref{eq:chihat} in \citet{mcdermott1988nonradial} has a typographical error.}
\begin{align}
\hat{\alpha}=\frac{1}{r^2}\frac{d}{dr}(r^2U)-\frac{\ell(\ell+1)}{r}V,
\label{eq:alphahat}
\end{align}
\begin{align}
\hat{\chi}=-\frac{\Gamma_1p}{\rho}\hat{\alpha}-\frac{1}{\rho}\frac{\partial p}{\partial r}U.
\label{eq:chihat}
\end{align}
\noindent In the crust these equations can be solved as a set of four first-order differential equations, whereas in the core they are simplified by the requirement that $\mu=0$, resulting in two first-order differential equations.

In this work we have applied Newtonian perturbations to a relativistic equilibrium stellar structure, resulting in a hybrid model. In order for this model to be usable, the modes must be orthogonal such that any perturbation can be expressed as a unique linear combination of the modes. The eigenfunction and eigenvalue of any mode can be defined in relation to the oscillation operator, $\mathcal{H}$, as shown by \citet{reisenegger1994multipole}:
\begin{align}
\mathcal{H}\xi=-\omega^2\xi.
\end{align}
\noindent Newtonian perturbations will only result in orthogonal modes if $\mathcal{H}$ is Hermitian with respect to the inner product of two vector fields (any two displacements of matter within the star), ie:
\begin{align}
\int_*\rho_0(r)\zeta^*(x)\cdot\mathcal{H}\Psi(x)d^3x=\int_*\rho_0(r)\mathcal{H}\zeta^*(x)\cdot\Psi(x)d^3x.
\end{align}
\noindent For a Newtonian stellar model the oscillation operator is Hermitian, and therefore applying Newtonian perturbations results in orthogonal modes. 

For a relativistic stellar model, the oscillation operator is not Hermitian. However, this does not pose a problem for a relativistic perturbation approach \citep[see e.g.][]{yoshida2002nonradial} because their eigenfrequencies are complex numbers, with the imaginary component arising from the damping of the mode due to the emission of gravitational waves. This imaginary component cancels out the deviation from orthogonality that arises from $\mathcal{H}$ not being Hermitian, and so relativistic perturbations can be applied to a relativistic stellar model to obtain orthogonal modes.

The hybrid model we have adopted can cause problems, because the stellar model does not give a Hermitian oscillation operator and the eigenfrequencies of Newtonian oscillations do not have the imaginary component required to cancel out the modes' deviation from orthogonality. To fix this, we follow \citet{reisenegger1994multipole} and define the local acceleration due to gravity within the star as
\begin{align}
g=-\frac{1}{\rho}\frac{dP}{dr}.
\label{eq:rel_grav}
\end{align}
\noindent This form for the gravity makes the oscillation operator Hermitian within the relativistic model, and therefore we can apply Newtonian perturbations to this modified relativistic star to obtain orthogonal modes.

At the boundary between the solid crust and the fluid core, there are three jump conditions that must be satisfied:\footnote{The version of the left-hand side of equation~\eqref{eq:jump2} in \citet{mcdermott1988nonradial} has a typographical error.}
\begin{align}
U\rvert_{r=R_{\rm cc}^{+}}=U\rvert_{r=R_{\rm cc}^{-}},
\label{eq:jump1}
\end{align}
\begin{align}
\frac{1}{p}\left(\lambda\hat{\alpha}+2\mu\frac{dU}{dr}\right)\biggr\rvert_{r=R_{\rm cc}^{+}}=\tilde{V}\left(\frac{U}{r}-\frac{\omega^2V}{g}\right)\biggr\rvert_{r=R_{\rm cc}^{-}},
\label{eq:jump2}
\end{align}
\begin{align}
\frac{\mu}{p}\left(\frac{dV}{dr}-\frac{V}{r}+\frac{U}{r}\right)\biggr\rvert_{r=R_{\rm cc}^{+}}=0,
\label{eq:jump3}
\end{align}
\noindent where 
\begin{align}
\tilde{V}=-\frac{d\ln\left(P\right)}{d\ln\left(r\right)}=\frac{\rho g r}{p},
\end{align}
\begin{align}
\lambda=\Gamma_1P-\frac{2}{3}\mu
\end{align}
\noindent is the Lam\'e coefficient, 
\begin{align}
M_r=4\pi\int^r_0 r'^2\rho(r')dr'
\label{eq:mass_r}
\end{align}
\noindent is the mass contained within radius $r$, and $r=R_{\rm cc}^{+}$ ($r=R_{\rm cc}^{-}$) indicates that the value is evaluated at the boundary when approached from the crust (core) of the star. The three conditions require different properties to be continuous across the crust-core boundary. The first is for the radial displacement, the second is for the pressure, and the third is for the transverse traction (which must be zero in the fluid core). The first two jump conditions can be combined to cancel out the arbitrary magnitude of $U$ and $V$ (which is different in the crust and the core), giving us
\begin{align}
\frac{r}{p}\frac{\left(\lambda\hat{\alpha}+2\mu\frac{dU}{dr}\right)}{U}\biggr\rvert_{r=R_{\rm cc}^{+}}=\tilde{V}\left(1-\frac{\sigma^2r}{g}\frac{V}{U}\right)\biggr\rvert_{r=R_{\rm cc}^{-}}.
\label{eq:jump2/1}
\end{align}
\noindent This leaves us with two jump conditions, equations~\eqref{eq:jump3} and~\eqref{eq:jump2/1}, and two eigenvalues, $\Omega$ and $V(R_*)$.

Every mode must satisfy the boundary conditions at the centre and surface of the star. The conditions at the surface are based on the requirements that the Lagrangian pressure perturbation and transverse traction go to zero at the surface:
\begin{align}
\frac{1}{p}\left(\lambda\hat{\alpha}+2\mu\frac{dU}{dr}\right)\biggr\rvert_{r=R_{*}}=\tilde{V}\left(\frac{\tilde{V}-c_1\Omega^2-4+\tilde{U}}{\frac{\ell(\ell+1)}{c_1\Omega^2}-\tilde{V}}+1\right)\frac{U}{r}\biggr\rvert_{r=R_{*}},
\label{eq:surface_boundary_modified_1}
\end{align}
\begin{align}
\frac{\mu}{p}\left(\frac{dV}{dr}-\frac{V}{r}+\frac{U}{r}\right)\biggr\rvert_{r=R_{*}}=0,
\label{eq:surface_boundary_modified_2}
\end{align}
\noindent where every quantity is evaluated at the surface of the star ($r=R_*$), and 
\begin{align}\nonumber
\tilde{U}=\frac{d\ln\left(M_r\right)}{d\ln\left(r\right)}=4\pi r^2\rho\frac{r}{M_r},\;\;\;c_1=\left(\frac{r}{R_*}\right)^3\frac{M_*}{M_r},\;\;\;\Omega^2=\frac{\omega^2R_*^3}{GM_*}
\end{align}
\noindent are equilibrium properties of the star. The condition at the centre of the star follows from the requirement that $U$ and $V$ be regular there:
\begin{align}
\left(\frac{c_1\Omega^2}{l}U-\frac{\sigma^2r}{g}V\right)\biggr\rvert_{r=0}=0,
\label{eq:core_condition}
\end{align}
\noindent where every quantity is evaluated at the centre of the star ($r=0$).

\subsection{The Crust-Core Interface Mode}
\begin{figure}
\centering
\includegraphics[width=0.45\textwidth,angle=0]{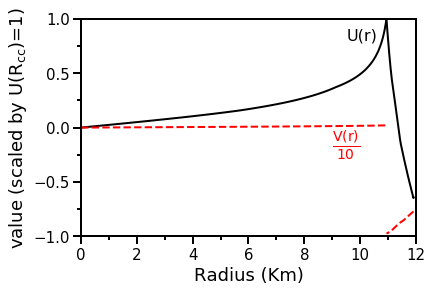}
\caption{The quadrupole crust-core interface mode for the EOS parameterised by $J=30$ MeV, $L=50$ MeV, and $K_{\rm sym}=-80$ MeV. $V(r)$ has been reduced by an order of magnitude so that it can be plotted alongside $U(r)$.}
\label{fig:2i_mode}
\end{figure}

We numerically solve for the eigenvalues by adjusting trial eigenvalues and solving equations~\eqref{eq:Ueqn} and~\eqref{eq:Veqn} until the jump conditions are satisfied, indicating that a mode has been found. 
For $J=30$ MeV, $L=50$ MeV, and $K_{\rm sym}=-80$ MeV the interface mode is found when the eigenvalues are $f=134.3$ Hz and $V(R_*)=-7.72$, resulting in the radial and transverse displacements shown in Figure \ref{fig:2i_mode}. This mode has a distinctive peak in radial displacement at the crust-core boundary, which is expected since the $i$-mode is caused by the discontinuity between the crust and core. The transverse displacement in the core is relatively small, with the discontinuity separating it from the larger displacement in the crust. Thus, a larger fraction of the mode energy goes into deforming the crust, helping it to reach the breaking strain faster. This makes the crust-core $i$-mode a good candidate to power an RSF.

\section{Results}
\subsection{The Impact of Symmetry Energy Parameters on Stellar Structure} \label{sec:stellar_structure}
\begin{figure}
\centering
\includegraphics[width=0.45\textwidth,angle=0]{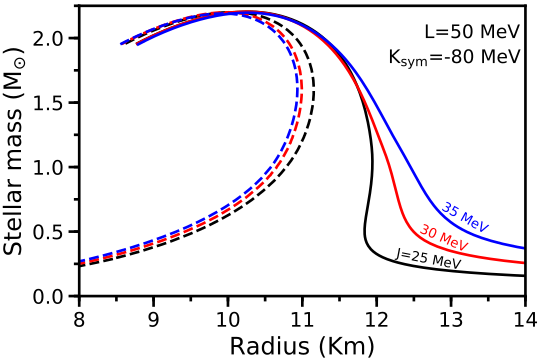}
\includegraphics[width=0.45\textwidth,angle=0]{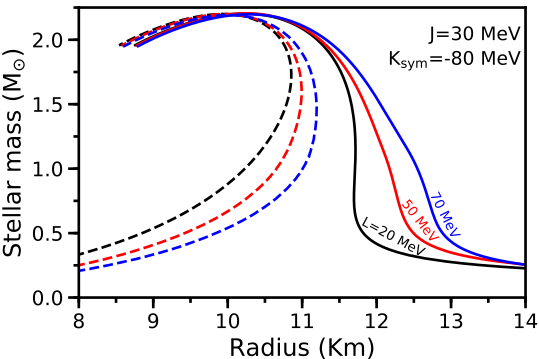}
\includegraphics[width=0.45\textwidth,angle=0]{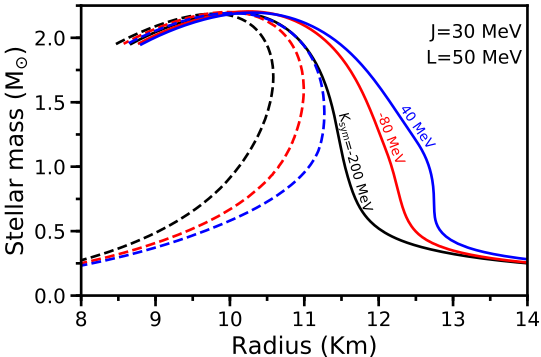}
\caption{The relationships between the neutron star mass and the stellar radius (solid lines) and the crust-core transition radius (dashed lines) for different EOSs. Each plot varies a different symmetry energy parameter over a wide range of values, with the lines being labelled with the varied parameter. The red (middle) lines of each plot are the same.}
\label{fig:MR_JLK}
\end{figure}

\begin{figure}
\centering
\includegraphics[width=0.45\textwidth,angle=0]{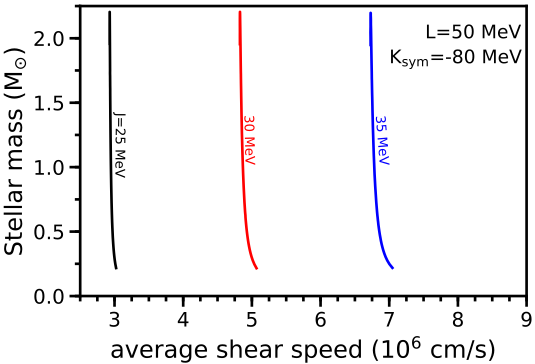}
\includegraphics[width=0.45\textwidth,angle=0]{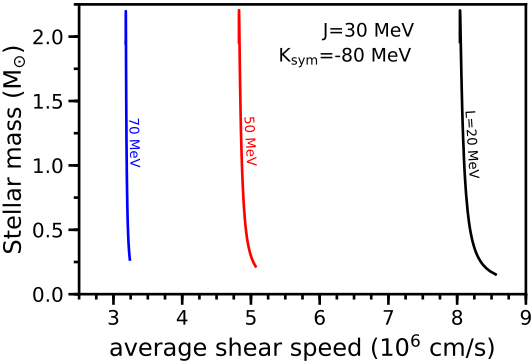}
\includegraphics[width=0.45\textwidth,angle=0]{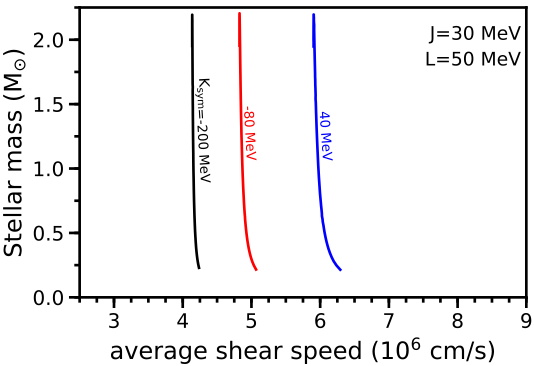}
\caption{The relationship between the neutron star mass and the density-weighted average of the shear speed for different EOSs. Each plot varies a different symmetry energy parameter over a wide range of values, with the lines being labelled with the varied parameter. The red (middle) lines of each plot are the same.}
\label{fig:Mct_JLK}
\end{figure}

Figure \ref{fig:MR_JLK} shows how the relationships between mass and stellar radius and between mass and crust-core transition radius change with the symmetry energy parameters $J$, $L$ and $K_{\rm sym}$. The maximum neutron star mass ($M_{\rm max}$) and moment of inertia at $1.4 M_\odot$ ($I_{1.4}$), which primarily control the core EOS, have been fixed as described in Section \ref{sec:core_model}. For many different types of mode, the frequency is dependant on both the stellar radius and the crust-core transition radius. Therefore, if this is the only thing we consider, Figure \ref{fig:MR_JLK} tells us that we would expect mode frequencies of $1.4$ M$_{\odot}$ neutron stars to vary the most with $K_{\rm sym}$ and the least with $J$. However, we must also consider the impact of the symmetry energy parameters on the restoring forces that cause the modes to oscillate. For the $i$-mode, the restoring forces are dominated by shear forces, and therefore we would expect the impact of the symmetry energy parameters on the $i$-mode frequency to be closely related to their impact on the shear speed, $c_t=\sqrt{\frac{\mu}{\rho}}$. Figure \ref{fig:Mct_JLK} shows how the relationship between the stellar mass and the density-weighted average of the shear speed in the crust changes with the symmetry energy parameters. This figure shows that $J$ and $L$ have larger impacts on the average shear speed than $K_{\rm sym}$, and that the shear speed is strongly dependent on all three of the symmetry energy parameters.

\begin{table}
\centering
\begin{tabular}{|c|c|c|c|c|}
\hline
Parameter&Change in&Change in&Change in\\
varied&$R_*$ (Km)&$R_{\rm cc}$ (Km)&$\bar{c}_t$ (cm/s)\\
\hline
\hline
$ $& $ $& $ $& $ $\\[-9pt]
$J$ ($25\rightarrow35$ MeV)&$0.21$&$-0.23$&$3.8\times 10^6$\\
\hline
$ $& $ $& $ $& $ $\\[-9pt]
$L$ ($20\rightarrow70$ MeV)&$0.47$&$0.48$&$-4.9\times 10^6$\\
\hline
$ $& $ $& $ $& $ $\\[-9pt]
$K_{\rm sym}$ ($-200\rightarrow40$ MeV)&$0.87$&$0.79$&$1.8\times 10^6$\\
\hline
\hline
$ $& $ $& $ $& $ $\\[-9pt]
$I_{1.4}$(I$_{\rm min}$ $\rightarrow$ I$_{\rm max}$)&$1.40$&$1.12$&$4.1\times 10^3$\\
\hline
\end{tabular}
\caption{Typical changes in stellar radius, crust-core transition radius, and density-weighted shear speed of a $1.4$ M$_{\odot}$ NS caused by varying each of the symmetry energy parameters. For each parameter, we vary it over the specified range while holding the others constant in the middle of their Uniform ranges. For comparison, the EOS in the middle of all of the Uniform ranges ($J=30$, $L=50$ and $K_{\rm sym}=-80$ MeV) results in $R_*=11.92$ Km, $R_{cc}=10.95$ Km, and $\bar{c}_t= 4.8\times 10^6$ cm/s. Also shown are the changes due to varying the moment of inertia of a $1.4$ M$_{\odot}$ NS (which determines the core EOS) across the full range of values allowed by causality while holding the three symmetry energy parameters constant and keeping the maximum mass fixed to M$_{\rm max}=2.2$ M$_{\odot}$.}
\label{tab:Parameter_vary}
\end{table}

In Table \ref{tab:Parameter_vary} we quantify the typical impact that varying the symmetry energy parameters has on the properties of a $1.4$ M$_{\odot}$ neutron star. From this table, and the trends of Figures \ref{fig:MR_JLK} and \ref{fig:Mct_JLK}, we see that varying the symmetry energy parameters causes a fractional change in the average shear speed that is significantly greater than the fractional change in the stellar radius or the transition radius. Therefore, we expect that the symmetry energy parameters' relationships to the $i$-mode frequency will be dominated by their relative contributions to the average shear speed. We will investigate this further in Section \ref{sec:shear_speed}, after we have calculated the dependence of the frequency on the symmetry energy parameters.

So far we have ignored the uncertainty in the parameters which control the core EOS. To address this, in Table \ref{tab:Parameter_vary} we also give the changes in stellar properties caused by varying the $1.4$ M$_{\odot}$ moment of inertia over an extremely conservative range. We find that this causes $\sim10$\% changes in the stellar radius and crust-core transition radius. These changes, while significant, are much smaller than the order unity changes in shear speed caused by varying the symmetry energy parameters. We therefore expect that, when compared to the symmetry energy parameters, the moment of inertia (and thus core EOS) has little impact on the $i$-mode frequency, and so we shall keep it fixed. The validity of this choice will be discussed further in Section~\ref{sec:discuss}. The maximum mass is more sensitive to the EOS at higher densities than the moment of inertia is, and so our results will be less sensitive to the choice of the maximum mass.

\subsection{Interface Mode dependence on Nuclear Parameters}
\label{sec:EOS_sets}
For the remainder of this paper, unless otherwise stated, we focus our results on $1.4M_{\odot}$ neutron stars. We explored the three different ranges of symmetry energy parameters described in Section \ref{sec:nuclear_model}. In Section \ref{subsec4_1_1}, we use our uniform (weakly-constrained) $J$, $L$ and $K_{\rm sym}$ ranges, in order to avoid tying our results to those of previous works. In Section \ref{subsec4_1_2} we use our PNM parameter ranges, where the parameters are consistent with the results of pure neutron matter calculations as this is the most relevant constraint for neutron star matter, which is extremely neutron rich. Finally, in Section \ref{subsec4_1_3} we use our MSL0-like parameter ranges, where $K_{\rm sym}$ is defined as a particular function of $J$ and $L$. This lets us more directly compare with previous works that have only allowed the first two symmetry energy parameters to vary, such as \citet{chen2010density}, \citet{steiner2012connecting} and \citet{tsang2009constraints}.

\subsubsection{Uniform (weakly-constrained) $J$, $L$ and $K_{\rm sym}$ ranges}\label{subsec4_1_1}
\begin{figure}
\centering
\includegraphics[width=0.45\textwidth,angle=0]{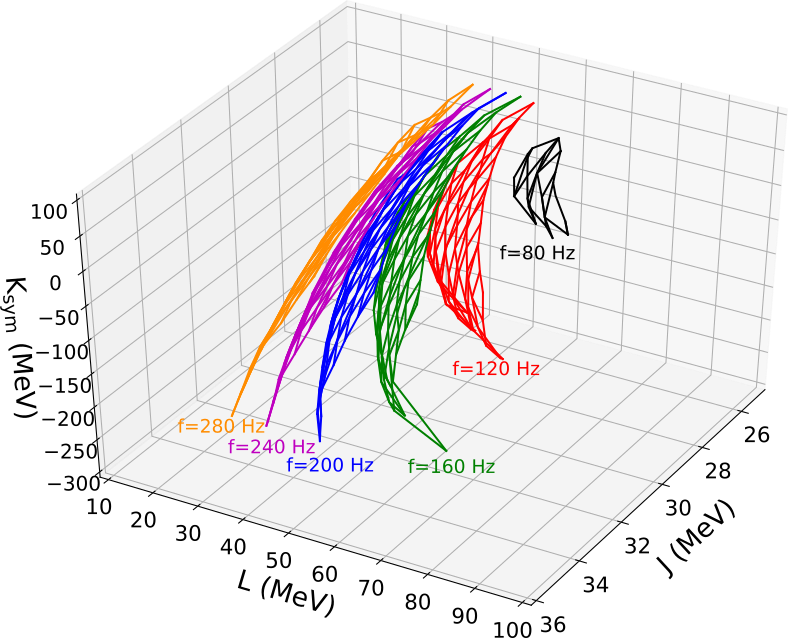}
\caption{Surfaces of constant $i$-mode frequency in the weakly-constrained symmetry energy parameter space of our uniform distribution. The surfaces are $40$ Hz apart. The data to manipulate this plot can be found at \url{https://github.com/davtsang/RSFSymmetry/}.}
\label{fig:grid_J_L_K}
\end{figure}

 We constructed a set of EOSs which had symmetry energy parameters evenly spaced in the three-dimensional parameter space defined for our uniform distribution in Section \ref{sec:nuclear_model} (we used a $J$ spacing of $1$ MeV, $L$ spacing of $10$ MeV, and $K_{\rm sym}$ spacing of $40$ MeV). After using the TOV equations to obtain a stellar model for each EOS, we calculated their $\ell = 2$ $i$-mode frequencies. We then interpolated between these frequencies to find surfaces of constant frequency in the $J$,$L$,$K_{\rm sym}$ parameter space, shown in Figure \ref{fig:grid_J_L_K}.

\begin{figure}
\centering
\includegraphics[width=0.45\textwidth,angle=0]{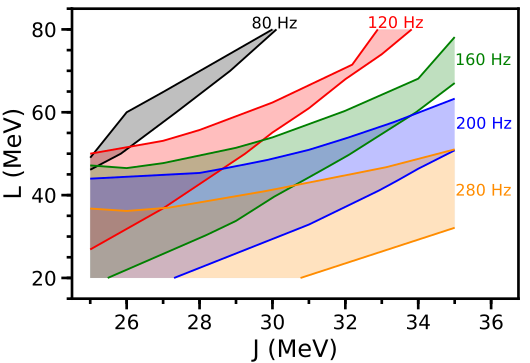}
\caption{The two-dimensional projection of Figure \ref{fig:grid_J_L_K} on the $J-L$ plane, showing the ranges of the symmetry energy parameters in which five example $i$-mode frequencies can be obtained. The widths of the regions are caused by $K_{\rm sym}$ having any value between $-200$ MeV and $40$ MeV.}
\label{fig:freq_contours_plane}
\end{figure}

In order to better understand Figure \ref{fig:grid_J_L_K} we plot its two-dimensional projection on the $J-L$ plane, shown in Figure \ref{fig:freq_contours_plane}. This figure shows the values of $J$ and $L$ that can result in the $i$-mode having the chosen frequency, with the spread in $L$ at any given $J$ being due to the range of possible $K_{\rm sym}$ values. We could also plot the projections on the $J-K_{\rm sym}$ and $L-K_{\rm sym}$ planes. However, we find that the strong dependence of the frequency on $L$ and $J$ means that these plots are uninformative, since the variation in the projected parameter can cause the $i$-mode frequency regions to cover almost the entire parameter space.

\begin{figure}
\centering
\includegraphics[width=0.45\textwidth,angle=0]{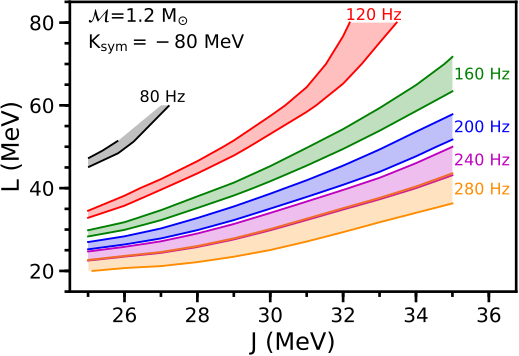}
\caption{The spread in the constraints on $J$ and $L$ caused by the resonance window (ie: the $i$-mode frequency can have any value between $f_{GW}-\delta f$ and $f_{GW}+\delta f$). We have fixed $K_{\rm sym}=-80$ MeV to isolate the effect of the resonance window.}
\label{fig:t_res_spread}
\end{figure}

One of the uncertainties affecting the constraints we could put on the symmetry energy parameters is the timescale over which resonant excitation of modes can occur. This can be calculated as \citep{tsang2012resonant}
\begin{align}
t_{\rm res}\sim 8\times10^{-2}\text{s}\left(\frac{\mathcal{M}}{1.2\text{M}_{\odot}}\right)^{\frac{-5}{6}}\left(\frac{f_{\rm mode}}{100 \text{ Hz}}\right)^{\frac{-11}{6}},
\label{eq:res_timescale}    
\end{align}
        \noindent where 
        \begin{align}
        \mathcal{M}=\frac{M_1^{\frac{3}{5}}M_2^{\frac{3}{5}}}{(M_1+M_2)^{\frac{1}{5}}}
        \label{eq:Mchirp}
        \end{align}
        \noindent is the chirp mass \citep{cutler1994gravitational}. 
This timescale can be combined with the rate of change of the gravitational wave frequency
\begin{align}
\Dot{f}_{\rm gw}=\frac{f_{\rm gw}}{4.7\times10^{-3}\text{s}}\left(\frac{\mathcal{M}}{1.2\text{M}_{\odot}}\right)^{\frac{5}{3}}\left(\frac{f_{\rm gw}}{1000 \text{ Hz}}\right)^{\frac{8}{3}}
\label{eq:df_GW/dt}    
\end{align}
\noindent (where $f_{\rm gw}\approx f_{\rm mode}$) to obtain a simple estimate of the range of frequencies over which resonance can occur:
\begin{align}
\delta f\sim t_{\rm res}\Dot{f}_{\rm gw}\sim3.7\text{Hz}\left(\frac{\mathcal{M}}{1.2\text{M}_{\odot}}\right)^{\frac{5}{6}}\left(\frac{f_{\rm mode}}{100 \text{ Hz}}\right)^{\frac{11}{6}}.
\label{eq:freq_spread}    
\end{align}
\noindent From this we see that the width of the resonance window increases with the frequency, with $\delta f$ scaling as $f_{\rm mode}^{\frac{11}{6}}$. For a chirp mass of $1.2M_{\odot}$ and a resonance at $100$ Hz, we get a frequency range of $\delta f\sim 3.7$ Hz, and for a resonance at $160$ Hz we get a range of $\delta f\sim 8.8$ Hz. This means that the spread of the frequency regions in the $J-L$ plane (seen in Figure \ref{fig:t_res_spread}) is quite small ($\delta L\lesssim 5$ MeV), and therefore the impact of the resonance window is significantly less than that of the $K_{\rm sym}$ range, which causes a spread of $\sim 20$ MeV in $L$. It should be noted that this is a very conservative estimate, and that by more accurately calculating both the rate at which energy is transferred into the modes and the breaking strain of the crust, this uncertainty could be significantly reduced by calculating the time it takes for the crust to shatter.

\begin{figure}
\centering
\includegraphics[width=0.45\textwidth,angle=0]{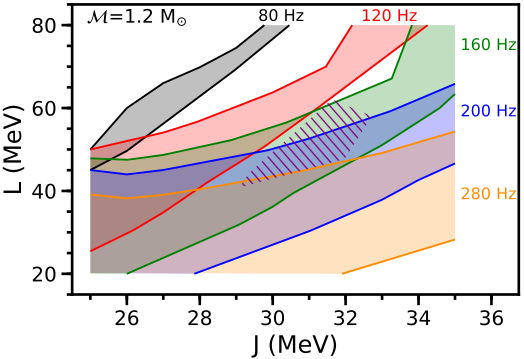}
\caption{Examples of constraints that could be applied to $J$ and $L$, with each constraint being labelled with the GW frequency at the time of the RSF. The width of the constraints comes from allowing $K_{\rm sym}$ to have any value in the range $-200$ to $40$ MeV at all $J$ and $L$ values, and from approximating the uncertainty in the $i$-mode frequency due to the resonance window with equation~\eqref{eq:freq_spread}. The hatched area in the centre of the plot represents the combined experimental nuclear constraints on $J$ and $L$ \citep{lattimer2013constraining}.}
\label{fig:all_constraints}
\end{figure}

In Figure \ref{fig:all_constraints} we plot regions in the $J-L$ plane that result in $i$-modes that can be resonantly excited by certain chosen GW frequencies when considering both the $K_{\rm sym}$ range (shown in Figure \ref{fig:freq_contours_plane}) and the resonance window (given by equation~\eqref{eq:freq_spread}). These regions are compared to the combined experimental nuclear constraints given in \citet{lattimer2013constraining} (which includes constrains from: fits to nuclear masses \citep{Kortelainen:2010aa}, neutron skin thickness \citep{chen2010density}, dipole polarisability \citep{Piekarewicz:2012aa}, giant dipole resonances \citep{trippa2008giant}, and isotope diffusion in heavy ion collisions \citep{tsang2009constraints}). From these results, in order to be consistent with the combined experimental nuclear constraints, we could expect to observe precursor flares in the range $120\lesssim f_{GW}\lesssim 280$ Hz. This range is very wide as it is based on our most conservative constraints on $J$ and $L$, with the upper bounds of Figure \ref{fig:all_constraints} using $f_{i} = f_{GW}-\delta f$ and $K_{\rm sym}=40$ MeV (or the maximum $K_{\rm sym}$ with a stable crust), and the lower bounds using $f_{i} = f_{GW}+\delta f$ and $K_{\rm sym}=-200$ MeV (or the minimum $K_{\rm sym}$ with a stable crust). In order to find more useful constraints on the symmetry energy parameters, we can reduce the $J$,$L$,$K{\rm sym}$ parameter space used to generate the EOSs.

\subsubsection{$J$, $L$ and $K_{\rm sym}$ constrained using pure neutron matter theory (PNM) } \label{subsec4_1_2}
\begin{figure}
\centering
\includegraphics[width=0.45\textwidth,angle=0]{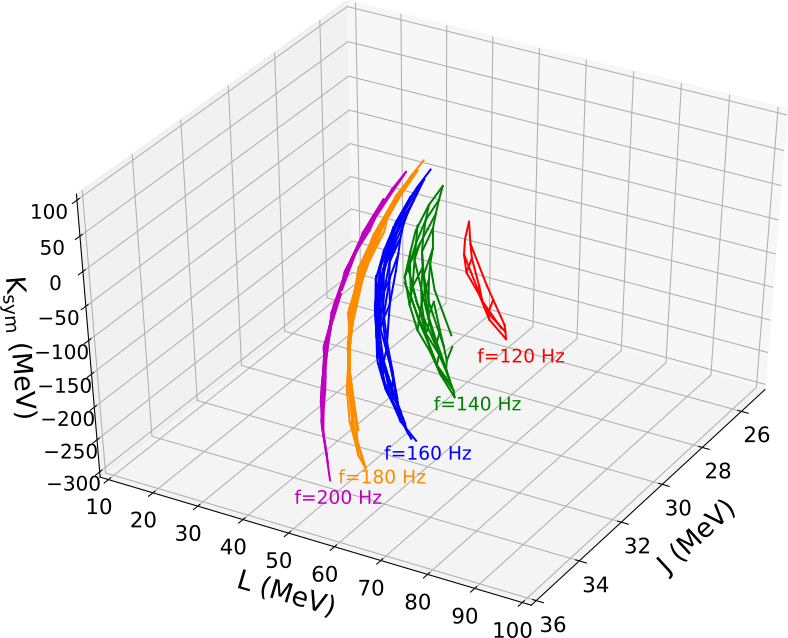}
\caption{Surfaces of constant $i$-mode frequency in the $J$,$L$,$K_{\rm sym}$ parameter space, where the $J$, $L$ and $K_{\rm sym}$ ranges are constrained by pure neutron matter theory. The surfaces are 20 Hz apart. The viewing angle is the same as for Figure \ref{fig:grid_J_L_K}, which causes the surfaces to collapse into lines. The data to manipulate this plot can be found at \url{https://github.com/davtsang/RSFSymmetry/}.}
\label{fig:PNM_planes}
\end{figure}

\noindent There are many different constraints on the nuclear symmetry energy parameters that we could use to reduce the $J$,$L$,$K_{\rm sym}$ parameter space. To keep our results conservative, we consider only the most relevant constraints so as to not make our results overly dependent on other works. For neutron star matter, which is extremely neutron rich, one such constraint comes from calculations of the properties of pure neutron matter (see Section \ref{sec:nuclear_model} and the PNM ranges discussed in \citet{newton2020nuclear}). With this additional constraint on the symmetry energy parameter ranges, we repeat our method from Section \ref{subsec4_1_1} for obtaining surfaces of constant frequency in the $J$,$L$,$K_{\rm sym}$ parameter space, resulting in Figure \ref{fig:PNM_planes}.

\begin{figure}
\centering
\includegraphics[width=0.45\textwidth,angle=0]{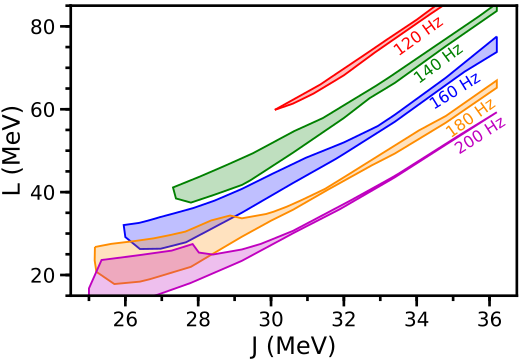}
\includegraphics[width=0.45\textwidth,angle=0]{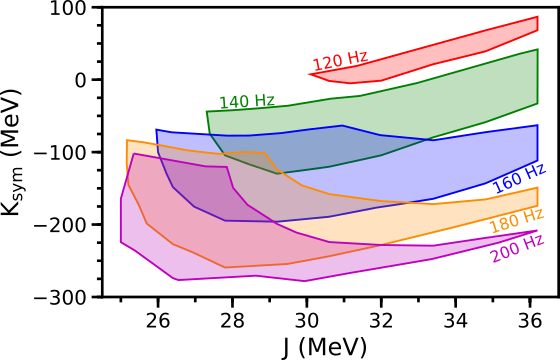}
\includegraphics[width=0.45\textwidth,angle=0]{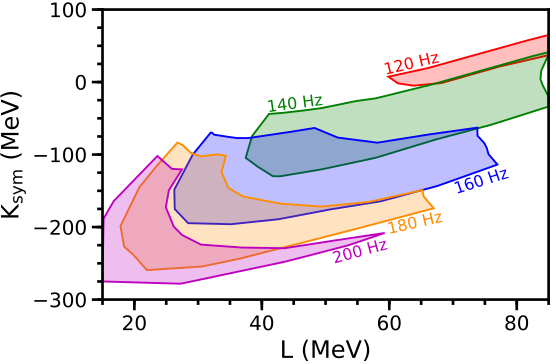}
\caption{The two-dimensional projections of Figure \ref{fig:PNM_planes}, showing the ranges of the symmetry energy parameters in which five example $i$-mode frequencies can be obtained while being consistent with pure neutron matter theory.}
\label{fig:PNM_2d}
\end{figure}

Figure \ref{fig:PNM_2d} shows two-dimensional projections of Figure \ref{fig:PNM_planes}. Similar to Figure \ref{fig:freq_contours_plane}, the first plot of this figure is the projection on the $J-L$ plane. However, 
by constraining the parameter space with the results of pure neutron matter theory we have reduced the range of $K_{\rm sym}$ values, and therefore the widths of the frequency regions are much smaller. Similarly, the ranges of $J$ and $L$ have been reduced, and therefore the projections of Figure \ref{fig:PNM_2d} on the $J-K_{\rm sym}$ and $L-K_{\rm sym}$ planes are now informative. These are shown in the second and third plots of Figure \ref{fig:PNM_2d}, where the widths of the frequency regions are determined by the ranges of $L$ and $J$ (respectively). In these three plots, the widths of the frequency regions show the impact of the projected symmetry energy parameter; the wider the regions, the more significant the uncertainty in the projected parameter is to the mode frequency.

\begin{figure}
\centering
\includegraphics[width=0.45\textwidth,angle=0]{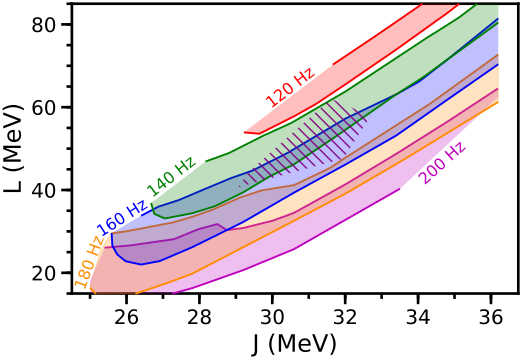}
\includegraphics[width=0.45\textwidth,angle=0]{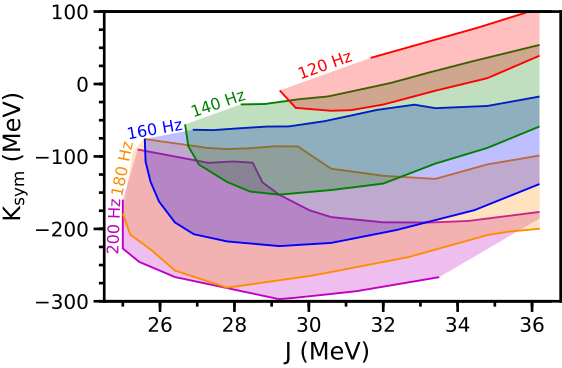}
\includegraphics[width=0.45\textwidth,angle=0]{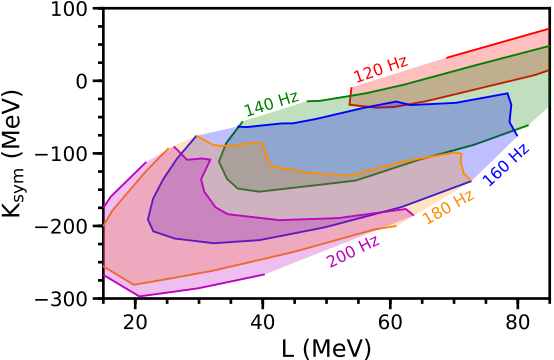}
\caption{The constraints that we could apply to the symmetry energy parameters in the event of an RSF detection at different GW frequencies, where the $J$, $L$ and $K_{\rm sym}$ ranges are constrained by pure neutron matter theory. The first (second) [third] plot shows the constraints on $J$ and $L$ ($J$ and $K_{\rm sym}$) [$L$ and $K_{\rm sym}$], with the strength of each constraint being determined by the $K_{\rm sym}$ ($L$) [$J$] range and the width of the resonance window. The hatched area in the centre of the first plot indicates the combined experimental nuclear constraints on $J$ and $L$.}
\label{fig:PNM_JL_res_spread}
\end{figure}

To calculate the constraints that we could place on the symmetry energy parameters, we combine the frequency regions shown in Figure \ref{fig:PNM_2d} with the uncertainty in the $i$-mode frequency due to the resonance window, given by equation~\eqref{eq:freq_spread}. This results in Figure \ref{fig:PNM_JL_res_spread}, which shows example constraints that could be applied to the symmetry energy parameters in the event of RSF detections at certain GW frequencies, alongside the combined experimental nuclear physics constraints on $J$ and $L$ from \citet{lattimer2013constraining}. By comparing Figure \ref{fig:all_constraints} and the first plot of Figure \ref{fig:PNM_JL_res_spread} we can see that restricting the symmetry energy parameters to the ranges predicted by pure neutron matter theory has significantly tightened our constraints, making them competitive with the experimental constraints. By inverting our method, we find that $120\lesssim f_{GW}\lesssim 180$ Hz results in constraints on $J$ and $L$ that are consistent with the combined experimental nuclear constraints.

\subsubsection{K$_{\rm sym}$ as a function of J and L (MSL0)}\label{subsec4_1_3}
\begin{figure}
\centering
\includegraphics[width=0.48\textwidth,angle=0]{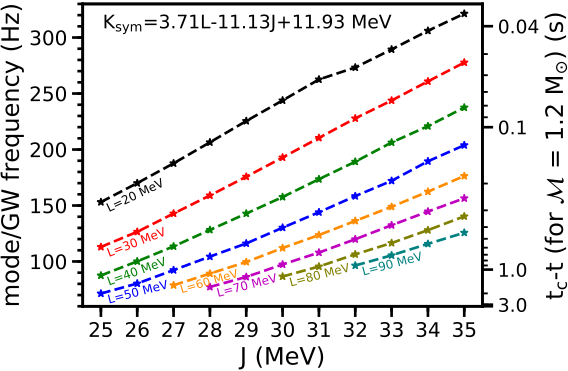}
\caption{The relationship between $J$, $L$ and the $i$-mode frequency for EOSs with the MSL0-like $K_{\rm sym}$ dependence. As there is no $K_{\rm sym}$ range, this plot shows every EOS used in Section \ref{subsec4_1_3}. The right axis of the plot shows the approximate times before coalescence at which the frequencies occur.}
\label{fig:MSL0_J_L_f_tc}
\end{figure}

To compare our work to more restricted two-parameter Skyrme models, we reproduce the MSL0 model's $K_{\rm sym}$ dependence on $J$ and $L$. Note that this dependence does not have any special physical significance, and there are other relationships between the symmetry energy parameters that are equally plausible. Using a similar grid of $J$ and $L$ values as in Section \ref{subsec4_1_1}, we calculated the $i$-mode frequencies for a set of stellar models to obtain Figure \ref{fig:MSL0_J_L_f_tc}. 
        This figure also shows the approximate relationship between frequency and time before coalescence, given by \citep{tsang2012resonant,blanchet2006gravitational}
        \begin{align}
        t_c-t=\frac{3}{8}t_{GW}=1.76\times 10^{-3}\rm{s}\left(\frac{\mathcal{M}}{1.2M_{\odot}}\right)^{-\frac{5}{3}}\left(\frac{f_{GW}}{1000 \rm{Hz}}\right)^{-\frac{8}{3}}.
        \label{eq:t_before_merger}
        \end{align}

\begin{figure}
\centering
\includegraphics[width=0.45\textwidth,angle=0]{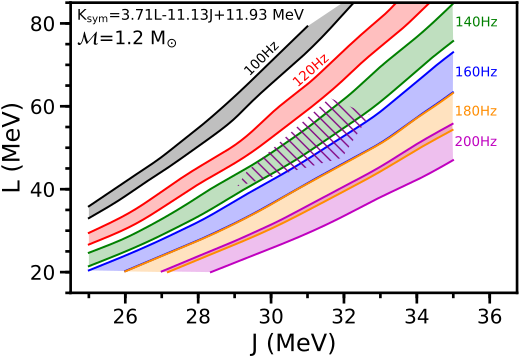}
\caption{Frequency contours in the $J-L$ plane that have been spread by the resonance window (with the MSL0-like $K_{\rm sym}$ dependence). The hatched area in the centre of the plot indicates the combined experimental nuclear constraints on $J$ and $L$.}
\label{fig:MSL0_shaded_constraints}
\end{figure}

We interpolated between the grid of $J$ and $L$ values to obtain frequency contours in the $J-L$ plane. These contours are spread by the resonance window calculated with equation~\eqref{eq:freq_spread}, resulting in the constraints shown in Figure \ref{fig:MSL0_shaded_constraints} (where we have also plotted the nuclear physics constraints on $J$ and $L$). These results represent a best case scenario for the $K_{\rm sym}$ range, as there is no uncertainty in its value at all $J$ and $L$ values. From this figure we can see that a precursor flare detected when $130\lesssim f_{GW}\lesssim 170$ Hz would provide constraints on $J$ and $L$ consistent with those from nuclear physics.

\begin{table}
\centering
\begin{tabular}{|c|c|c|}
\hline
$J$, $L$, $K_{\rm sym}$&GW frequency&Time before\\
 ranges&of the RSF (Hz)&coalescence (s)\\

\hline
$ $& $ $& $ $\\[-9pt]
Uniform&$120$-$280$&$0.50$-$0.05$\\
\hline
$ $& $ $& $ $\\[-9pt]
PNM&$120$-$180$&$0.50$-$0.17$\\
\hline
$ $& $ $& $ $\\[-9pt]
MSL0&$130$-$170$&$0.41$-$0.20$\\
\hline
\end{tabular}
\caption{Summary of, for each of our data sets, the approximate range of gravitational wave frequencies (and corresponding times before coalescence, assuming a chirp mass of $1.2$ $M_{\odot}$) in which a Resonant Shattering Flare needs to occur in order to be consistent with the combined experimental nuclear constraints on $J$ and $L$ given in \citet{lattimer2013constraining}. These results are for the Skyrme model described in \citet{newton2020nuclear}, using Newtonian perturbations and a $M_*=1.4$ M${_\odot}$ neutron star. They do not include the impact of the core parameterisation, which is fixed as described in Section \ref{sec:core_model}.}
\label{tab:freq_to_match_nuclear}
\end{table}

Table \ref{tab:freq_to_match_nuclear} inverts our method for constraining the symmetry energy parameters by showing the approximate range of gravitational wave frequency in which we would expect to observe an RSF in order for our constraints on $J$ and $L$ to be consistent with the combined experimental constraints \citep{lattimer2013constraining} (ie: they have a non-zero overlap). From this we can see that, for the model used in this work and a $1.4$ M$_{\odot}$ neutron star, we expect to observe RSFs at gravitational wave frequencies of around $150$ Hz, or approximately $0.3$ s before coalescence. This is similar to the time before the main SGRB that many precursors are observed ($~0.1-5.0$ s), providing evidence that these precursors are RSFs.

\subsection{Shear speed}\label{sec:shear_speed}

\begin{figure}
\centering
\includegraphics[width=0.45\textwidth,angle=0]{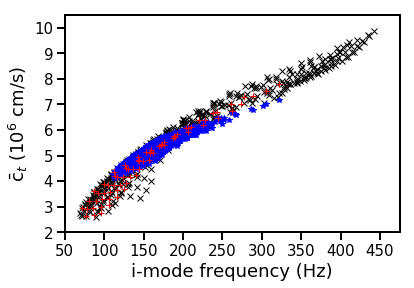}
\caption{The relationship between the $i$-mode frequency and the density-weighted average shear speed for each of our EOSs. The black markers are for EOSs in our Uniform set, the blue for EOSs in our PNM set, and the red for EOSs in our MSL0 set.}
\label{fig:f_avct_allEOSs}
\end{figure}

In order to determine the cause of the change in $i$-mode frequency due to variations in $J$, $L$ and $K_{\rm sym}$, we investigated how the properties of the star we discussed in Section \ref{sec:stellar_structure} relate to the frequency. As we predicted, the frequency was strongly dependent on the density-weighted average of the shear speed in the crust. This is shown in Figure \ref{fig:f_avct_allEOSs}, which relates the frequency and the average shear speed for our three sets of EOSs.

\begin{figure}
\centering
\includegraphics[width=0.45\textwidth,angle=0]{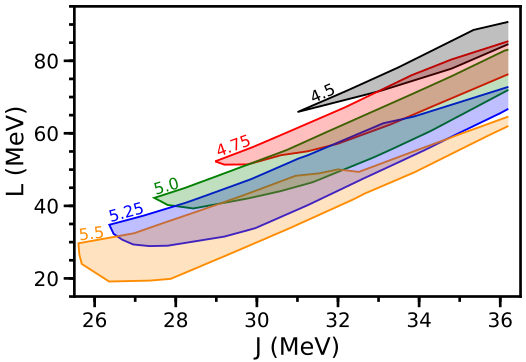}
\caption{Similar to the first plot of Figure \ref{fig:PNM_2d}, but for the density-weighted average shear speed. The spread is due to the $K_{\rm sym}$ range, which is consistent with pure neutron matter theory, and the regions are labelled in $10^6 \rm{cm/s}$.}
\label{fig:avct_JL_PNM}
\end{figure}

In a similar way to how the first plot of Figure \ref{fig:PNM_2d} shows the $J$ and $L$ values that can result in the $i$-mode having chosen frequencies, Figure \ref{fig:avct_JL_PNM} shows the $J$ and $L$ values that can result in stars with chosen average shear speeds. The similarities between the regions shown in this figure and in Figure \ref{fig:PNM_2d} indicate that the $i$-mode frequency and average shear speed are closely linked. At higher $J$ and $L$ values these figures become less similar, suggesting that the significance of other stellar properties increases with $J$ and $L$. In this figure we have only shown the results for our PNM set of EOSs, since all three sets of EOSs give the same qualitative results.

\section{Discussion} \label{sec:discuss}

For all three sets of EOSs, our constraints on $J$ and $L$ (shown in Figures \ref{fig:all_constraints}, \ref{fig:PNM_JL_res_spread} and \ref{fig:MSL0_shaded_constraints}) are angled in the same direction as the combined constraints from other works \citep{lattimer2013constraining}. Therefore, a detection of an RSF at a frequency in the middle of the range that is consistent with these constraints would provide a small improvement to our knowledge of $J$ and $L$. However, if an RSF were to be detected at a higher or lower frequency our constraints could be more interesting due to their overlap with the combined experimental nuclear constraints being smaller.

\begin{figure}
\centering
\includegraphics[width=0.45\textwidth,angle=0]{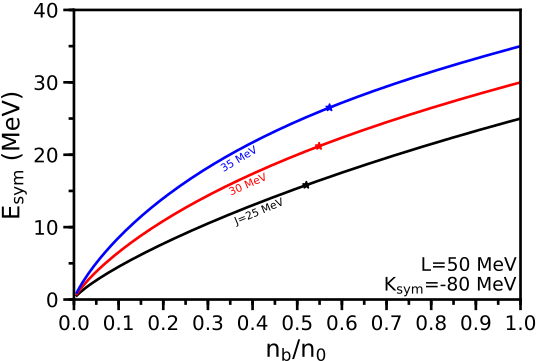}
\includegraphics[width=0.45\textwidth,angle=0]{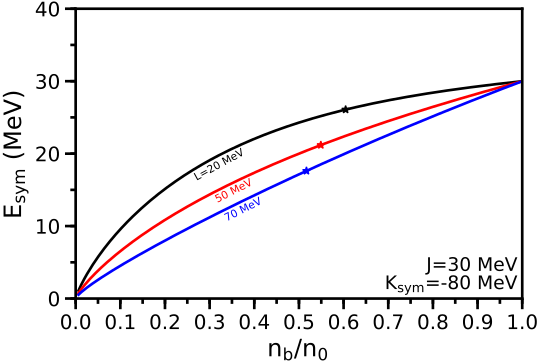}
\includegraphics[width=0.45\textwidth,angle=0]{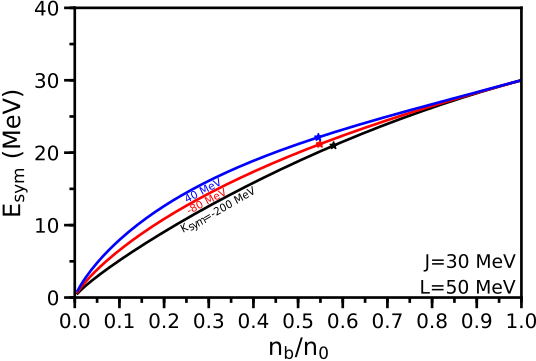}
\caption{The symmetry energy below nuclear saturation density for different combinations of symmetry energy parameters. Each plot varies a different parameter over a wide range of values, with the lines being labelled with the varied parameter. The markers indicate the crust-core transition density, showing how it varies with the symmetry energy parameters. The red (middle) lines of each plot are the same.}
\label{fig:Esym_JLK}
\end{figure}

The shear speed increasing with $J$ and $K_{\rm sym}$ and decreasing with $L$ is correlated with the impact of these changes on the symmetry energy in the crust (where $n_{\rm b}<n_0$); increasing $J$ and $K_{\rm sym}$ and decreasing $L$ causes the symmetry energy at crustal densities to increase, as can be seen in Figure \ref{fig:Esym_JLK}. Here the markers indicate the crust-core transition, although note that the crust-core transition density does not strongly correlate with the crust thickness - see e.g. \citep{Margueron:2011}. Increasing the symmetry energy increases the energy cost of creating more neutrons, and therefore decreases the fraction of dripped neutrons in the crust. Equation~\eqref{eq:neutrons_ions} shows that as the dripped neutron fraction decreases, the ion number density increases (for a fixed mass number of nucleus in the crust). This leads to an increase in the shear modulus, as can be seen in equation~\eqref{eq:mu_1991_a} or~\eqref{eq:altmu}. By calculating the density-averaged neutron fraction $\bar{X}_{\rm n}$, mass number $\bar{A}$ and charge number $\bar{Z}$ we have confirmed that changes in $\bar{X}_{\rm n}$ are the dominant outcome of varying any of the three symmetry energy parameters.

Together, Figures \ref{fig:Esym_JLK} and \ref{fig:f_avct_allEOSs} provide a qualitative physical understanding of the $i$-mode frequency dependence shown in Figures \ref{fig:all_constraints}, \ref{fig:PNM_JL_res_spread}, and \ref{fig:MSL0_shaded_constraints}. The symmetry energy profile of the star determines the composition and shear modulus in the inner crust, on which the $i$-mode frequency is strongly dependent.

It is important to note that equation~\eqref{eq:mu_1991_a} is a fit to calculations of the shear modulus for an ionic lattice with no dripped neutrons and ionic separations typical of the outer crust. The shear modulus of the deep inner crust [including the nuclear pasta layers \citep{Pethick:1998aa}] remains an important outstanding problem, the result of which might significantly affect our results. However, the fact that the shear modulus depends on the ion separation, which in turn depends on the fraction of dripped neutrons, means that the relationship between the $i$-mode frequency and the symmetry energy is likely to persist.

\begin{figure}
\centering
\includegraphics[width=0.45\textwidth,angle=0]{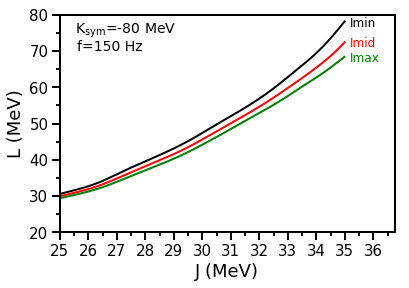}
\caption{The effect of changing I$_{1.4}$ (which is used to parameterise the core EOS) on the symmetry energy parameters that give a frequency of $150$ Hz, where we have fixed $K_{\rm sym}=-80$ MeV to isolate the impact of the core EOS. I$_{\rm min}$ is the minimum moment of inertia allowed by causality, I$_{\rm max}$ is the maximum, and I$_{\rm mid}$ is their average. For the set of EOSs plotted here, I$_{\rm min}$ ranges from 1.03$\times 10^{45}$ g cm$^2$ to 1.2$\times 10^{45}$ g cm$^2$ and I$_{\rm max}$ ranges from 1.44$\times 10^{45}$ g cm$^2$ to 1.57$\times 10^{45}$ g cm$^2$. The moment of inertia increases with $J$ and $L$.}
\label{fig:I1.4}
\end{figure}

To test that our results are not significantly affected by the choice of core EOS, we can investigate the impact of allowing the moment of inertia parameter (I$_{1.4}$) to vary between the minimum and maximum values allowed by causality. Having a range of I$_{1.4}$ values does not noticeably affect the shear speed, but it does increase the range of stellar radii obtained with our sets of EOSs. In Table \ref{tab:Parameter_vary} we show the typical changes in relevant stellar properties caused by increasing I$_{1.4}$ from its minimum to maximum value while holding the symmetry energy parameters constant. If the symmetry energy parameters are all allowed to vary in their Uniform ranges, and I$_{1.4}$ is held constant at the average of the minimum and maximum values allowed by causality, the radius of a $1.4$ M$_{\odot}$ NS ranges from $10.6$ to $12.7$ Km. If we also allow I$_{1.4}$ to vary between its minimum and maximum values, the radius ranges from $9.90$ to $13.6$ Km.Both of these radius ranges are obtained while assuming that the NS maximum mass is $2.2$ M$_{\odot}$. From Figure \ref{fig:I1.4} we can see that the impact of the core EOS on the $i$-mode frequency is negligible at low $L$ and $J$, and is still small at higher values. This is because the change in radius caused by the core EOS is far less significant than the change in shear speed caused by the symmetry energy parameters. This illustrates that resonant shattering flares mainly probe the EOS and composition of the neutron star crust, in contrast to tidal deformability measurements that give information about the core EOS.

We can quantify the impact of the core EOS by calculating the change in $i$-mode frequency caused by varying I$_{1.4}$. For all $J$, $L$ and $K_{\rm sym}$ values in our `uniform' ranges, compared to the average value of I$_{1.4}$ we find that the maximum and minimum I$_{1.4}$ cause approximately $-5\%$ and $+6\%$ changes in $i$-mode frequency (respectively). As the I$_{1.4}$ range used here is only constrained by causality, it is extremely conservative and therefore the choice of core EOS does not significantly affect our results. We also note that the same event that results in a coincident detection of an RSF can be used to extract the tidal deformability. This parameter constrains the core EOS in a complimentary manner to the constraints explored in this work, with the added restriction that neutron star masses and EOSs are the same in the description of each phenomenon.

\begin{figure}
\centering
\includegraphics[width=0.45\textwidth,angle=0]{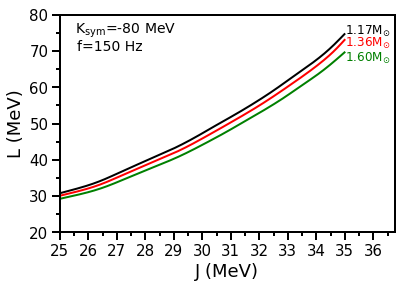}
\caption{The effect of changing the total NS mass on the symmetry energy parameters that give a frequency of $150$ Hz, where we have fixed $K_{\rm sym}=-80$ MeV to isolate the impact of the stellar mass. These mass values are the 90\% confidence ranges calculated by \citet{abbott2017gw170817} for the stars in GW170817 (using low-spin priors).}
\label{fig:vary_mass_contours}
\end{figure}

We have assumed a neutron star mass of $1.4$ M$_{\odot}$, but from Figure \ref{fig:vary_mass_contours} we can see that a realistic degree of uncertainty in the stellar mass \citep{abbott2017gw170817} has a noticeable impact on the symmetry energy parameters that give a chosen $i$-mode frequency. The change in the chosen frequency contour is similar to the impact of the core EOS shown in Figure \ref{fig:I1.4}. However, the moment of inertia range used in Figure \ref{fig:I1.4} is very conservative, and so the uncertainty in the neutron star mass measurement is likely to have a more significant impact on the symmetry energy parameter constraints than the uncertainty in the core EOS. Uncertainty in the mass of an RSF's source should be considered when calculating constraints on the symmetry energy parameters, as its impact is similar to that of the resonance window (as can be seen by comparing Figures \ref{fig:vary_mass_contours} and \ref{fig:t_res_spread}).

In this work we have used the hybrid approach of non-relativistic perturbations of a relativistic star to obtain the wave-equation (ignoring dynamic perturbations of the gravitational potential). Using relativistic perturbations (while still ignoring metric perturbations) as in \citet{yoshida2002nonradial} can result in $\sim10$\% changes in the mode frequencies. This is significant when compared to the width of our constraints on $J$ and $L$, and we will explore this effect on the constraints in a future work.

Accurately calculating the Schwarzschild discriminant is not simple, and as it has little impact on the $i$-mode we have set it to zero (i.e. we have assumed the star is barotropic). We have also assumed that the binary lifetime is much longer than the neutron stars' spin-down and cooling timescales, and so we have ignored rotational and high temperature effects. Finally, we have not considered the impact of superfluidity in the core of a neutron star. Superfluidity allows protons and neutrons to move somewhat independently of each-other, introducing a new set of counter-moving normal modes \citep{andersson2001dynamics}, as well as modifying the frequency of modes that mainly oscillate within the core of the star. In the inner crust, partial entrainment of the superfluid may change the shear speed by reducing the effective density that accelerates due to shear forces.

\section{Conclusion}
We have calculated the relationship between the neutron star interface mode frequency and the first three parameters that characterise the density dependence of the nuclear symmetry energy at saturation density ($J$, $L$ and $K_{\rm sym}$). This was done by using an extended Skyrme mean-field model for the crust and outer core of the star, supplemented by two polytropes that controlled the high density EOS. We have used this to present potential constraints on the symmetry energy parameters that could be obtained by coincident multimessenger detection of a Resonant Shattering Flare and gravitational wave chirp during a binary neutron star inspiral. These constraints have been shown to be competitive with current nuclear experimental constraints.

Previous works have shown \citep{abbott2017gw170817,bauswein2017neutron,abbott2018gw170817,de2018tidal,abbott2019properties} that the gravitational wave chirp from a binary neutron star inspiral (with sufficient signal-to-noise ratio) can constrain the tidal deformabilities, masses, and radii of the stars. These, in turn, place constraints on the neutron star equation of state \citep{Read:2009ab,read2013matter,lackey2015reconstructing,margalit2017constraining,annala2018gravitational,lim2018neutron,most2018new,Fattoyev:2018aa,carson2019constraining,zhang2019extracting,landry2019Nonparametric,essick2020nonparametric}, primarily in the core. In this work, we have examined the nuclear physics constraints (in particular on the nuclear symmetry energy parameters $J$, $L$, and $K_{\rm sym}$) that could be obtained by a future detection of a Resonant Shattering Flare along with a gravitational-wave chirp. Timing of the RSF relative to the GW chirp can provide a direct measurement of the resonant frequency of the $\ell = 2$ core-crust interface mode \citep{tsang2012resonant}. This frequency is dependent on properties of the neutron star near the core-crust boundary, and is thus sensitive to the nuclear symmetry energy parameters which determine (in a model dependent way), the properties of the neutron star near nuclear saturation. The measurement of an $i$-mode frequency through coincident timing of an RSF would provide astrophyiscal constraints orthogonal to those sensitive mainly to the core EOS.

Following \citet{newton2020nuclear}, we constructed three sets of EOSs parameterised by $J$, $L$ and $K_{\rm sym}$, with each set allowing these parameters to have different ranges. The high density EOS parameters were fixed by choosing a reasonable value for $M_{\rm max}$ and a representative value of the moment of inertia of a 1.4M$_{\rm odot}$ star, $I_{1.4}$. Solving for the $i$-mode frequencies, we were able to determine the region in the parameter space to which $J$, $L$ and $K_{\rm sym}$ could be constrained given measurements of different frequency values.

Multimessenger coincident timing of an RSF would give the $i$-mode frequency to a precision roughly determined by the duration of the flare. Additionally, taking the conservative assumption that the nuclear symmetry energy parameters are consistent with the results of pure neutron matter theory provides constraints on $J$, $L$, and $K_{\rm sym}$ that are competitive with \citet{Kortelainen:2010aa}, \citet{chen2010density} and \citet{tsang2009constraints}. Conversely, we can use the constraints found by other works to obtain the range of frequencies in which we would expect to observe an RSF for a $1.4$ M$_{\odot}$ neutron star. For the models used in this work, the range predicted by pure neutron matter theory is $\sim 120-180$ Hz.

We have shown that it is important to take into account the variation of the third symmetry energy parameter, $K_{\rm sym}$, independent of $J$ and $L$. For example, if we allow all three to vary, our predicted range of $i$-mode frequencies is $120$-$280$ Hz, while if $K_{\rm sym}$ is restricted by a choice of model, an artificially smaller range is predicted ($130$-$170$ MeV in the case of the MSL0 model considered here). Conversely, experimental measurements of $K_{\rm sym}$ will constrain the predicted range of frequencies.

In Figure \ref{fig:f_avct_allEOSs} we showed that the $i$-mode frequency (a global property of the NS) is strongly dependent on the average (density-weighted) shear speed within the crust (a local material property of the crust). Therefore, the dependence of the frequency on the symmetry energy parameters is dominated by their effects on the shear modulus within the crust, and in particular near the crust-core boundary. Figures \ref{fig:Mct_JLK} and \ref{fig:avct_JL_PNM} related the shear speed to the symmetry energy parameters \citep[similar to Figure 1 of ][]{steiner2009constraints}, connecting changes in these nuclear physics parameters to their impact on the average shear speed in the crust. While other global properties of the stellar structure (e.g. neutron star radius, radius of the core-crust transition) which vary with the model parameters ($J$, $L$, $K_{\rm sym}$, $M_{\rm max}$ and $I_{1.4}$) also play a role, we found that the $i$-mode frequency depends most strongly on the average shear speed, as can be seen from the similarities between Figure \ref{fig:avct_JL_PNM} and Figure \ref{fig:PNM_JL_res_spread}.

The quantitative results presented in this work are model dependant. Our focus is on constraining the symmetry energy parameters that characterize the crust and out core EOS, so that is where we span the widest range of the available parameter space by independently varying the first three parameters in the Taylor expansion around saturation density. However, we do restrict the parameter space of $J$ and $L$ to that spanned by nuclear experimental constraints; notably, this means values of $L$ are, for the most part, below 90 MeV. This excludes some of the stiffest EOSs, and therefore the neutron star models with the largest possible radii. A recent measurement of the neutron skin of $^{208}$Pb suggests that the slope of the symmetry energy $L$ may be significantly above 100 MeV \citep{2021arXiv210103193R}, which, although at odds with most other experimental results, is a reminder we should not rule out stiffer EOSs. The parameter space of the high density EOS, consisting of two polytropes, is restricted to a maximum neutron star mass of $2.2$ M$_{\odot}$, and a moment of inertia of a 1.4$M_{\odot}$ star in the middle of the range allowed by causality. Using different EOS models may result in significantly different frequencies, with \citet{tsang2012resonant} showing $i$-mode frequencies as low as $30$ Hz. However, we have investigated the impact of variation in the moment of inertia parameter and found that it had little impact on the $i$-mode, as it did not significantly affect the shear speed. Therefore, when choosing models for use in the analysis of RSFs, the description of the neutron star crust is the most important input. A exploration of the wider parameter space including high-$L$ EOSs will be the subject of future work.

A number of upcoming nuclear experiments promise to constrain the symmetry energy further. We highlight the ongoing efforts to extract the neutron skin of neutron rich nuclei from measurements of the parity-violating asymmetry in the electron scattering cross-section caused by the weak interaction \citep{Abrahamyan:2012aa} at Jefferson Lab and Mainz Superconducting Accelerator \citep{Horowitz:2014aa,Becker:2018aa,Thiel:2019aa}. The latter is responsible for the recent measurement of the neutron skin mentioned above. As illustrated in figure 1 of \citet{steiner2009constraints}, neutron skins provide a constraint on the symmetry energy that is orthogonal to those provided by the constraints on the shear speed and hence the $i$-mode frequency. Powerful constraints may be obtained in the future by combining these weak, EM and gravitational-wave observations to probe the strong force in multi-messenger nuclear astrophysics.

Using upcoming LIGO/Virgo/KAGRA observing runs \citep{abbott2020prospects}, and existing Gamma-ray burst monitors such as Swift/BAT \citep{barthelmy2005burst} and Fermi/GBM \citep{meegan2009fermi} to provide coincident timing, the detection of a Resonant Shattering Flare during a binary neutron star inspiral can provide a new complementary astrophysical constraint on nuclear physics parameters by probing the bulk properties of neutron star matter near the crust/core transition. The rates of RSFs are currently uncertain, with precursor flares estimated to occur for $\sim3-10$\% of SGRBs. However, the recent coincident detection of an (off-axis) SGRB and the chirp from GW170817 suggests a rate of NS mergers such that we may soon be able to obtain these powerful constraints.

\section*{Acknowledgements}
DN is supported by a University Research Studentship Allowance from the University of Bath, and WGN acknowledges support from NASA grant 80NSSC18K1019. We would like to thank the anonymous referee whose insightful suggestions have greatly improved the clarity and presentation of this work. 


\section*{Data Availability}
The code to calculate the stellar models and $i$-mode frequencies, along with the tabulated EOSs and compositions used for the grids of symmetry energy parameters are provided via \url{https://github.com/davtsang/RSFSymmetry/}.



\bibliographystyle{mnras}
\bibliography{RSFSymmetry,newton} 




%
%


\bsp	
\label{lastpage}
\end{document}